\documentclass[
    aps,
    prb,
    twocolumn,
    superscriptaddress,
    amsmath,
    amssymb
]{revtex4-2}

\usepackage{braket}
\usepackage{graphicx}
\usepackage[table]{xcolor}
\usepackage{siunitx}
\usepackage{gensymb}
\usepackage{bm}
\usepackage{multirow}
\usepackage{bbm}
\usepackage{enumitem}
\usepackage{oplotsymbl}
\usepackage{bbding}
\usepackage{orcidlink}
\usepackage{pifont}
\usepackage{array}
\usepackage{booktabs}
\usepackage{float}
\usepackage{hyperref}

\hypersetup{
    colorlinks=true,
    allcolors=blue
}

\newcommand{\dd}{\mathrm{d}}

\begin{document}

%----------------------------------------------------------
\title{Dirac-like bands and weak-field transverse transport in ABA trilayer graphene}

\author{Agustin Montemurro\,\orcidlink{0009-0006-4281-1217}}
\thanks{These authors contributed equally to this work.}
\affiliation{Department of Physics and Astronomy, University of Manchester,
Oxford Road, Manchester, M13 9PL, United Kingdom}
\affiliation{National Graphene Institute, University of Manchester,
Booth St.\ E., Manchester, M13 9PL, United Kingdom}

\author{Fujia Lu\,\orcidlink{0009-0004-2115-9309}}
\thanks{These authors contributed equally to this work.}
\affiliation{Department of Physics and Astronomy, University of Manchester,
Oxford Road, Manchester, M13 9PL, United Kingdom}
\affiliation{National Graphene Institute, University of Manchester,
Booth St.\ E., Manchester, M13 9PL, United Kingdom}

\author{Ellis R. Huntley\,\orcidlink{0009-0002-5396-9897}}
\affiliation{Department of Physics and Astronomy, University of Manchester,
Oxford Road, Manchester, M13 9PL, United Kingdom}
\affiliation{National Graphene Institute, University of Manchester,
Booth St.\ E., Manchester, M13 9PL, United Kingdom}

\author{Vladimir Fal'ko\,\orcidlink{0000-0003-0828-0310}}
\email{vladimir.falko@manchester.ac.uk}
\affiliation{Department of Physics and Astronomy, University of Manchester,
Oxford Road, Manchester, M13 9PL, United Kingdom}
\affiliation{National Graphene Institute, University of Manchester,
Booth St.\ E., Manchester, M13 9PL, United Kingdom}

% \date{}
%----------------------------------------------------------
\begin{abstract}
The low-energy electronic structure of ABA trilayer graphene is composed of light and heavy electron branches, which can be altered by an out-of-plane displacement field. Here, we show that, depending on the values of Slonczewski-Weiss-McClure parameters, a displacement field can inflict a band edge inversion, for which we identify the inversion conditions. This inversion is accompanied by a reversal of Berry curvature and giant topological magnetic moments at the band edges. In a weak perpendicular magnetic field, those topological magnetic moments produce valley splitting, giving rise to distinct anomalous transverse electrical and thermoelectric responses. These features offer a potential route to detecting the band inversion and constraining several of the least established tight-binding parameters, such as the on-site energy shift for the dimer pair of carbons $\Delta^\prime$ and the next neighbour hopping parameters $\gamma_2$ and $\gamma_5$.
\end{abstract}
%----------------------------------------------------------
\maketitle
%----------------------------------------------------------
\section{Introduction}\label{sec:introduction}

\begin{figure}[t]
    \centering  \includegraphics[width=0.95\linewidth]{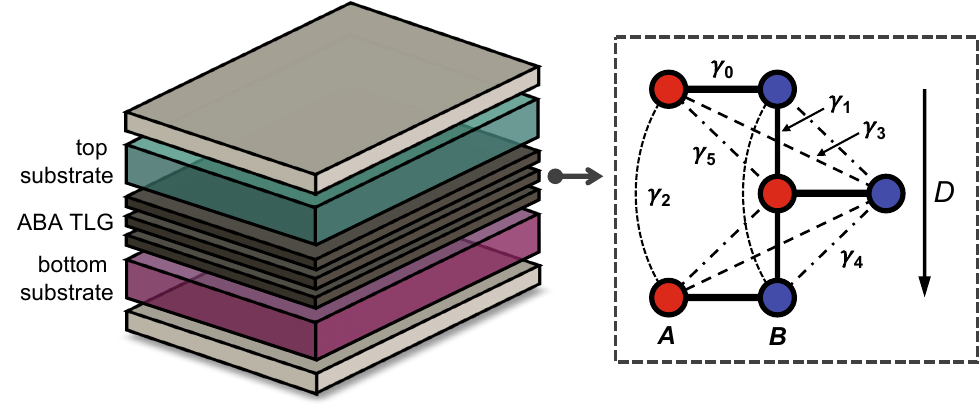}
    \caption{ Schematic of the electrostatic setup for controlling the carrier density $n$ and displacement field $D$, and of a side-view of ABA-TLG with the labelled tight-binding model parameters.}
    \label{fig:setup}
\end{figure}

The electronic and topological properties of few-layer graphene depend sensitively on the number of layers and their stacking order. Monolayer graphene exhibits a linear low-energy dispersion near the Dirac points \cite{Novoselov2005,Zhang2005,CastroNeto2009}, whereas Bernal-stacked bilayer graphene features approximately quadratic low-energy bands with an electrically tunable gap \cite{McCannFalko2006,Zhang2009,Knothe2018}. Adding a third layer introduces two distinct stacking configurations, rhombohedral (ABC) and Bernal (ABA), with qualitatively different low-energy electronic structures \cite{bao_stacking-dependent_2011,slizovskiy_films_2019}. 
This structural diversity makes trilayer graphene a versatile platform for exploring a range of electronic phenomena, including gate-tunable semimetallic behaviour \cite{craciun_trilayer_2009,henriksen_quantum_2012}, unconventional quantum Hall states \cite{taychatanapat_quantum_2011,zibrov_emergent_2018,winterer_spontaneous_2022}, giant orbital magnetic moments and valley splitting \cite{ge_control_2021}, as well as ordinary and nonlinear Nernst responses \cite{Rezaei2024,Liu2025,Varshney2026}. Recent studies of magic-angle twisted trilayer graphene have also revealed the coexistence and interplay of light and heavy electrons \cite{Pierce2025}. Within the Bernal-stacked family, ABA trilayer graphene (TLG) occupies a particularly distinct position. 
While a monolayer constitutes the fundamental graphene building block and an AB bilayer forms the natural two-layer unit cell of Bernal graphite, an ABA trilayer represents an odd-layer structure rather than a complete repetition of the AB unit \cite{Koshino2010,Malard2009}, as illustrated in Fig.~\ref{fig:setup}. It therefore contains a unique central layer and, when the two outer layers experience equivalent electrostatic environments, possesses a mirror symmetry about that layer \cite{bao_stacking-dependent_2011, koshino_gate-induced_2009, koshino_landau_2011}.

This mirror symmetry provides a natural classification of the electronic states. Symmetric and antisymmetric combinations of the outer-layer orbitals have opposite mirror parity and cannot hybridise. The low-energy six-band Hamiltonian separates into a mirror-even, bilayer-like branch and a mirror-odd, monolayer-like branch. ABA TLG is therefore the minimal Bernal-stacked system in which monolayer-like and bilayer-like quasiparticles coexist within the same material \cite{taychatanapat_quantum_2011}.

Unlike ABC TLG, pristine ABA TLG does not possess inversion symmetry. Furthermore, the inequivalent dimer and non-dimer environments of the central and outer layers, represented by the on-site energy parameter ($\Delta^\prime$) and the remote Slonczewski–Weiss–McClure (SWMcC) hopping parameters, open gaps within the individual low-energy subbands. The complete system may nevertheless remain semimetallic because the monolayer-like and bilayer-like bands overlap in energy \cite{craciun_trilayer_2009, henriksen_quantum_2012}. These subband gaps generate valley-contrasting Berry curvature and orbital magnetic moment even in the absence of an externally imposed asymmetry between
the outer layers \cite{xiao_valley-contrasting_2007,xiao_berry_2010,ge_control_2021}.

An out-of-plane displacement field provides a means of tuning the low-energy electronic structure of ABA TLG by inducing an electrostatic asymmetry between the outer layers and breaking mirror symmetry \cite{koshino_gate-induced_2009,morimoto_gate-induced_2013}. Previous studies of ABA TLG have shown a wide range of estimates for SWMcC parameters, particularly the on-site energy difference $\Delta'$ and the remote hopping parameters $\gamma_2$ and $\gamma_5$, which determine the low-energy Dirac-like band \cite{Zhou2023,Shi2020}. These discrepancies make it important to examine different parameter combinations in theoretical studies. Notably, different choices of these parameters can lead not only to quantitative changes in the band gaps but also to qualitatively different band-edge ordering. This raises the possibility of a displacement-field-induced inversion of the Dirac-like band edges, depending on the tight-binding parameters \cite{morimoto_gate-induced_2013,Mann2025}.

In this paper, starting from a six-band Hamiltonian with self-consistent electrostatic screening, we study the low-energy electronic structure of ABA TLG under an out-of-plane displacement field.
We examine how the displacement field modifies the coupling between the monolayer-like and bilayer-like branches, and derive an analytical condition for the inversion of the Dirac-like band edges, depending on the tight-binding parameters. 
Then we analyse the associated topological properties, including the reversal of Berry curvature near the valley centre, changes in valley Chern numbers, and the enhancement of orbital magnetic moments near the band edges.
To provide a more comprehensive picture of the transverse transport properties, we calculate both the anomalous and conventional contributions to the Hall and transverse thermoelectric conductivities under a weak perpendicular magnetic field, and examine their relative importance in the total response.

%----------------------------------------------------------
\section{Model}\label{sec:model}

We consider a dual-gated ABA TLG heterostructure encapsulated by top and bottom dielectric substrates, as illustrated in Fig.~\ref{fig:setup}. ABA TLG is the minimal Bernal-stacked graphene system whose low-energy description requires the full SWMcC parameterisation \cite{mcclure_band_1957,slonczewski_band_1958}. We therefore retain the six hopping parameters $\gamma_i$ ($i=0,\ldots,5$), together with the stacking-induced on-site energy $\Delta^\prime$, as illustrated in Fig.~\ref{fig:setup}. The values of these parameters extracted from experiments and first-principles calculations are known with some uncertainty, depending on the measurement and simulation techniques \cite{Zhou2023}. Here, we use the values $\gamma_0=3.16$~eV, $\gamma_1=0.39$~eV, $\gamma_2=-0.02$~eV, $\gamma_3=0.316$~eV, $\gamma_4=0.044$~eV and $\gamma_5=0.038$~eV. The electrostatic environment enters through the outer-layer potentials $\Delta_1$ and $\Delta_2$, which are determined by the externally applied displacement field, dielectric environments, and self-consistent Hartree screening.

In the basis of sublattice Bloch states
\{
$\psi_{A_1}$, 
$\psi_{B_1}$, 
$\psi_{A_2}$, 
$\psi_{B_2}$, 
$\psi_{A_3}$, 
$\psi_{B_3}$
\}, 
where $A_i$ and $B_i$ denote the two sublattice
sites in layer $i$, the low-energy Hamiltonian is \cite{koshino_gate-induced_2009}
\begin{align} \label{Hamiltonian}
   & H ({\bf k}) = \notag\\
   &\begin{pmatrix}
  	\Delta_1 & v \pi^\dagger & -v_4 \pi^\dagger & v_3 \pi & \gamma_2/2 & 0 \\
  	v \pi & \Delta' + \Delta_1 & \gamma_1 & -v_4 \pi^\dagger & 0 & \gamma_5/2 \\
  	-v_4 \pi & \gamma_1 & 2 \Delta' & v \pi^\dagger & -v_4 \pi & \gamma_1 \\
  	v_3 \pi^\dagger & -v_4 \pi & v \pi & 0 & v_3 \pi^\dagger & -v_4 \pi \\
  	\gamma_2/2 & 0 & -v_4 \pi^\dagger & v_3 \pi & \Delta_2 & v \pi^\dagger \\
  	0 & \gamma_5/2 & \gamma_1 & -v_4 \pi^\dagger & v \pi & \Delta' + \Delta_2 \\
    \end{pmatrix}.
\end{align}
Here $\pi=\hbar(\xi k_x+ik_y)$, where $\xi=\pm1$ label the valleys $\mathbf K$ and $\mathbf K'$, and $\mathbf{k}$ is measured from the corresponding valley centre. The velocities are defined by $\hbar v=\sqrt{3}a\gamma_0/2$ and $\hbar v_i=\sqrt{3}a\gamma_i/2$ for $i=3,4$, with lattice constant $a=0.246\,\mathrm{nm}$. We take $e>0$, so the electron charge is $-e$.
The electrostatic environment is described by the layer potentials $\Delta_1$ and $\Delta_2$, which represent the on-site energy shifts of the bottom and top graphene layers relative to the middle layer, respectively. 
It is convenient to decompose the outer-layer potentials into their symmetric and antisymmetric components,
\begin{equation}
    \Delta_s = \frac{\Delta_1 + \Delta_2}{2},
    \qquad
    \Delta_a = \frac{\Delta_2-\Delta_1}{2}.
\label{eq:layer_potentials}   
\end{equation}
The symmetric component $\Delta_s$ preserves mirror symmetry with respect to the middle layer, whereas a finite $\Delta_a$ breaks this symmetry. These parameters and notations will be used throughout the rest of the manuscript when analysing possible regimes of trilayer band structure.

\subsection{Mirror Symmetric ABA Trilayer Graphene}
\label{subsec:mirror sym}
Mirror reflection about the central layer allows the electronic states to be classified by their parity. Using symmetric and antisymmetric combinations of the outer-layer orbitals $\psi_{A_\pm} = (\psi_{A_1}\pm\psi_{A_3})/\sqrt{2}$ and $\psi_{B_\pm} = (\psi_{B_1}\pm\psi_{B_3})/\sqrt{2}$,
the Hamiltonian separates into a mirror-odd monolayer-like branch and a mirror-even bilayer-like branch, $H(\Delta_a=0)=H_m\oplus H_b$, with

\begin{equation}
\begin{aligned}
H_m &=
\begin{pmatrix}
-\dfrac{\gamma_2}{2}+\Delta_s & v\pi^\dagger \\
v\pi & -\dfrac{\gamma_5}{2}+\Delta'+\Delta_s
\end{pmatrix},
\\[10pt]
H_b &=
\begin{pmatrix}
\dfrac{\gamma_2}{2}+\Delta_s
& v\pi^\dagger
& -\sqrt{2}v_4\pi^\dagger
& \sqrt{2}v_3\pi \\
v\pi
& \dfrac{\gamma_5}{2}+\Delta'+\Delta_s
& \sqrt{2}\gamma_1
& -\sqrt{2}v_4\pi^\dagger \\
-\sqrt{2}v_4\pi
& \sqrt{2}\gamma_1
& 2\Delta'
& v\pi^\dagger \\
\sqrt{2}v_3\pi^\dagger
& -\sqrt{2}v_4\pi
& v\pi
& 0 \nonumber
\end{pmatrix}.
\end{aligned}
\label{eq:mirror_blocks}
\end{equation}

Here, $H_m$ and $H_b$ represent the $2\times2$ mirror-odd (monolayer-like) and $4\times4$ mirror-even (bilayer-like) Hamiltonian blocks, respectively.
This decomposition exposes the two low-energy components of ABA TLG. The mirror-odd block has the Dirac-like dispersion inherited from monolayer graphene, whereas the mirror-even block contains the quadratic low-energy bands associated with Bernal bilayer graphene.

Although $\Delta_s$ shifts the monolayer-like bands, it does not affect their energy gap, as it enters both diagonal elements of $H_m$ equally. The monolayer-like branch is therefore not inherently massless, with an asymmetry between $A_{-}$ and $B_{-}$ sites controlled by the following combination of SWMcC parameters: $\Delta^\prime+(\gamma_2-\gamma_5)/2$.
 
A finite antisymmetric layer potential $\Delta_a$ breaks mirror symmetry and introduces coupling between the monolayer-like and bilayer-like branches. Therefore, the full Hamiltonian takes the form
\begin{equation}
\begin{aligned}
   H &=
   \begin{pmatrix}
   H_m & V_a\\
   V_a^\dagger & H_b
   \end{pmatrix}; \ \    V_a = 
	\begin{pmatrix}
        -\Delta_a & 0 & 0 & 0 \\
		0 & -\Delta_a & 0 & 0  
	\end{pmatrix}. \nonumber
\end{aligned}    
\end{equation}
Below we relate those asymmetry parameters to the external displacement field by performing their self-consistent electrostatic (Hartree) analysis.

%----------------------------------------------------------
\subsection{Self-Consistent Hartree Screening}

An external perpendicular displacement field $\mathbf{D}=D\hat{\mathbf{z}}$, controlled by the top- and bottom-gate voltages $V_{\mathrm{tg}}$ and $V_{\mathrm{bg}}$, induces a potential difference across the trilayer. Interlayer charge redistribution screens this applied field through the self-consistent Hartree potential. Including the dielectric environment and Hartree screening, the resulting potential shifts on the bottom and top layers, $\Delta_1$ and $\Delta_2$, and the layer-resolved carrier densities $n_i$ are given by \cite{Garcia-Ruiz2023}
\begin{align}
\Delta_1 &= \frac{eDd}{\epsilon_0 \epsilon_z} + \Delta_b - \Delta_{1h}, \nonumber \\
\Delta_2 &= -\frac{eDd}{\epsilon_0 \epsilon_z} + \Delta_t + \Delta_{2h}, \\
    n_i &= 4\int\frac{\dd^2k}{(2\pi)^2}\sum^6_{\beta=1}\bigg[\big(|\psi^{A_i}_\beta|^2 + |\psi^{B_i}_\beta|^2\big) f(\varepsilon_\beta - \mu) -\frac{1}{6}\bigg]. \notag
\end{align}   
Here, the terms $\Delta_{b}$ and $\Delta_{t}$ denote the electrostatic offsets associated with the bottom and top dielectric environments, respectively, while $\Delta_{1h}$ and $\Delta_{2h}$ are the Hartree screening corrections.
In the expression for $n_i$, $\beta$ labels the bands, $A_i$ and $B_i$ label the two sublattices belonging to layer $i$, and the prefactor of 4 accounts for spin and valley degeneracies. The corresponding Hartree corrections obtained from the Poisson equation for the layered geometry are 
\begin{equation}
\begin{aligned}
\Delta_{1h}&=
\frac{e^2d}{2\epsilon_0}
\left[
(n_2-n_1)\frac{1+\epsilon_z}{2\epsilon_z}
+\frac{n_3}{\epsilon_z}
\right], \\ 
\Delta_{2h}&=
\frac{e^2d}{2\epsilon_0}
\left[
(n_3-n_2)\frac{1+\epsilon_z}{2\epsilon_z}
-\frac{n_1}{\epsilon_z}
\right],
\end{aligned}
\end{equation}
where $\epsilon_z$ is the effective out-of-plane permittivity between the graphene layers. The layer potentials are solved self-consistently together with the electronic occupations for each carrier density and external displacement field.

The self-consistent screening results in a non-zero $\Delta_s$, even in the absence of an externally imposed layer asymmetry \cite{koshino_gate-induced_2009}. Within the SWMcC description, the different local environments of the middle- and outer-layer dimer sites, parametrised in particular by $\Delta^\prime$, generate an intrinsic interlayer charge redistribution and hence a finite symmetric potential $\Delta_s$.

\begin{figure}[t]
    \centering  \includegraphics[width=0.95\linewidth]{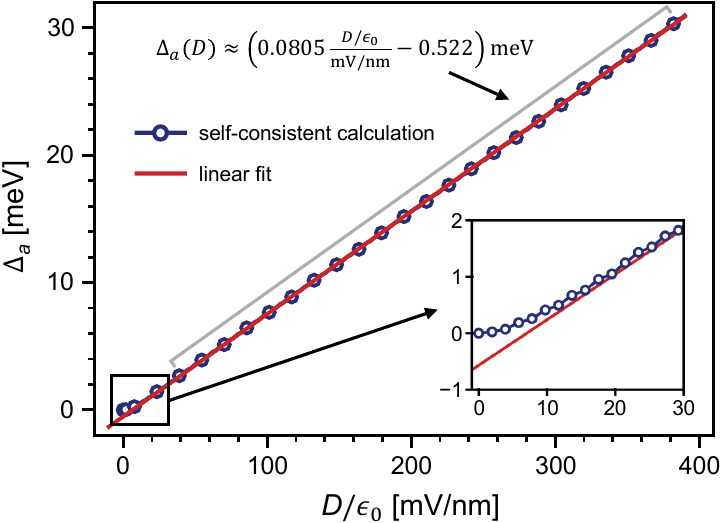}   
    \caption{Self-consistent screening of the external displacement field in ABA TLG. The antisymmetric layer potential $\Delta_a$ is shown as a function of the external displacement field $D/\epsilon_0$ calculated at charge neutrality for $\Delta^\prime=35$~meV. The blue line with open circles represents the self-consistent Hartree result, while the red line shows a linear fit. The linear relationship for large displacement fields provides an empirical screening coefficient for relating the antisymmetric potential to the realised displacement field.}
    \label{fig:screening}
\end{figure}

On the other hand, such screening determines the antisymmetric potential $\Delta_a$ for a given displacement field $D$. 
For $\Delta'=35$ meV at charge neutrality, the antisymmetric layer potential $\Delta_a$ initially exhibits a nonlinear increase with the displacement field $D$, before rapidly approaching an approximately linear dependence, as shown in Fig.~\ref{fig:screening}.
By fitting the high-field regime of the numerical calculations [blue line with open circles in Fig.~\ref{fig:screening}], we write
$\Delta_a(D) \approx \left(0.0805\frac{D/\epsilon_0}{\mathrm{mV/nm}}-0.522\right)~\text{meV}$ 
[see red line in Fig.~\ref{fig:screening}].
The resulting $\Delta_a(D)$ incorporates the effects of self-consistent screening and controls the coupling between the monolayer- and bilayer-like branches. Varying $D$ field therefore tunes their hybridisation and the resulting low-energy band structure.

%----------------------------------------------------------
\subsection{Quantum Topological Quantities}

The geometric properties of the electronic bands are characterised by their Berry curvature.
For each band $\beta$, the Berry curvature is defined as $\Omega_{\beta}(\mathbf{k})=
i\langle \boldsymbol{\nabla}_{\mathbf{k}}u_\beta |
\times
|\boldsymbol{\nabla}_{\mathbf{k}}u_\beta\rangle
\cdot\hat{\mathbf{z}}$, where $|u_\beta\rangle$ is the periodic part of the Bloch function \cite{xiao_berry_2010,xiao_valley-contrasting_2007}.
In numerical calculations, we evaluate the Berry curvature using the equivalent Kubo formula, 
\begin{equation}
\Omega_{\beta}^{\xi}(\mathbf{k})
=
-2\,\mathrm{Im}
\sum_{\beta'\neq\beta}
\frac{
\langle u_\beta^\xi|\partial_{k_x}H(\mathbf{k})|u_{\beta'}^\xi
\rangle
\langle u_{\beta'}^\xi|\partial_{k_y}H(\mathbf{k})|u_\beta^\xi
\rangle
}{
(\varepsilon_\beta^\xi-\varepsilon_{\beta'}^\xi)^2
},
\label{eq:berry_kubo}
\end{equation}
where $H(\mathbf{k})$ is described by Eq. \eqref{Hamiltonian}, and $\varepsilon_\beta$ are energies of its eigenstates $|u_{\beta}\rangle$. This formula avoids taking momentum derivatives of the Bloch eigenstates and is therefore independent of their gauge choice. Near band degeneracies, the small energy denominators enhance the Berry curvature, whose redistribution between the bands accompanies the closing and reopening of the gap.

For an isolated band, the Berry curvature integrated over the momentum-space region $S_\xi$ associated with valley $\xi$ defines the valley Chern number, 
$\mathcal{C}_\beta^\xi=
(1/2\pi)
\int_{S_\xi}\dd^2k\,
\Omega_{\beta}^\xi(\mathbf{k})$. 
Time-reversal symmetry (which relates $\bf K$ valley [$\xi =+1$] with $\bf K^\prime$ valley [$\xi=-1$]) requires that the Berry curvature satisfies $\Omega_{\beta}^{\xi}(\mathbf{k}) = -\Omega_{\beta}^{-\xi}(-\mathbf{k})$.

\subsection{Quantum Topological Contributions to Electrical and Thermoelectric Transport}

Below we analyse both classical Hall and transverse thermoelectric responses as well as contributions from Berry-curvature-induced drift of electrons \cite{xiao_berry_2010}, 
$$\dot{\mathbf r}_{\beta,\xi}=\hbar^{-1}\nabla_{\mathbf k}\varepsilon^\xi_\beta(\mathbf k)-\dot{\mathbf k} \times {\bf \hat z} {\Omega}^\xi_\beta(\mathbf k),$$  towards both effects.

At zero magnetic field, the anomalous Hall contributions from the two valleys cancel because the corresponding states have the same energy and occupation. However, a weak perpendicular magnetic field can lift this cancellation through its coupling to the orbital magnetic moment
$$
    m_{z,\beta}(\mathbf{k})=
    -\frac{ie}{2\hbar}
    \langle \boldsymbol{\nabla}_{\mathbf{k}}u_\beta | \times [H(\mathbf{k})
    -\varepsilon_\beta(\mathbf{k})]
    |\boldsymbol{\nabla}_{\mathbf{k}}u_\beta\rangle
    \cdot\hat{\mathbf{z}}.
$$ 
The orbital magnetic moment is odd under time reversal, with $m_{z,\beta}^{\xi}(\mathbf{k})=-m_{z,\beta}^{-\xi}(-\mathbf{k})$. As a result, the magnetic field shifts the band energies according to $\tilde{\varepsilon}_\beta^\xi(\mathbf{k})=\varepsilon_\beta^\xi(\mathbf{k})-m_{z,\beta}^\xi(\mathbf{k})B_z$, producing opposite energy shifts in the two valleys. These valley-dependent energy shifts modify the electronic occupations and prevent the anomalous Hall contributions from cancelling completely.
Consequently, we adopt a linear expansion
$f(\tilde{\varepsilon})\approx f(\varepsilon)-m_z B_z (\partial f/\partial \varepsilon)$ 
and write the corresponding anomalous Hall conductivity as
\begin{equation}
\sigma^{\mathrm A}_{xy} = \frac{2e^2B_z}{\hbar}\sum_{\beta,\xi}\int_{S_\xi}\frac{\dd^2k}{(2\pi)^2}
  \frac{\partial f}{\partial\varepsilon}\,m^\xi_{z,\beta}(\mathbf k)\,\Omega^\xi_{z,\beta}(\mathbf k).
\label{eq:ahe_linear_B}
\end{equation}

In contrast, the conventional Hall response arises from the Lorentz-force correction to the semiclassical distribution. Within the constant relaxation time approximation, its linear-in-$B_z$ contribution is

\begin{equation}
\begin{split}
\sigma^{\mathrm{cl}}_{xy} &= -\frac{e^3\tau^2B_z}{2\pi^2\hbar^4}\sum_{\beta,\xi}\int \dd^2k\,
  \frac{\partial f}{\partial\varepsilon^\xi_\beta}\,\mathcal F^\xi_\beta(\mathbf k), \\
\mathcal F_\beta^\xi(\mathbf k)
&=
\frac{\partial\varepsilon_\beta^\xi}{\partial k_x}
\frac{\partial\varepsilon_\beta^\xi}{\partial k_y}
\frac{\partial^2\varepsilon_\beta^\xi}
{\partial k_x\partial k_y}-
\left(
\frac{\partial\varepsilon_\beta^\xi}{\partial k_x}
\right)^2
\frac{\partial^2\varepsilon_\beta^\xi}{\partial k_y^2}.
\end{split}
\label{eq:conventional_hall}
\end{equation}
We can see that the conventional response depends on the Fermi-surface velocities, band curvature, and the relaxation time $\tau$, whereas the anomalous contribution is determined by the geometric factor $m_z\Omega_z$ and is independent of $\tau$ within this treatment.
The total Hall conductivity is then given by $\sigma_{xy}^{\text H}=\sigma_{xy}^{\text{cl}}+\sigma_{xy}^{\text{A}}$.

In addition to the electrical response, a temperature gradient can also drive an electric current. The linear transport response is described by the electrical and
thermoelectric conductivity tensors,
\begin{equation}
    j_i = \sum_j \left(
    \sigma_{ij}E_j-\beta_{ij}\nabla_j T
    \right), \nonumber
\end{equation}
where $j_i$ is the electric current density, $E_j$ is the electric field, and $\nabla_j T$ is the temperature gradient. Here, $\sigma_{ij}$ and $\beta_{ij}$ denote the electrical and thermoelectric conductivity tensors, respectively.

The corresponding transverse thermoelectric conductivities are obtained by
weighting the same transport kernels by the energy carried relative to the
chemical potential \cite{xiao2006prl}. This gives
\begin{subequations}
\begin{eqnarray}
\beta^{\mathrm{cl}}_{xy} &=& \frac{e^2\tau^2B_z}{2\pi^2\hbar^4T}\sum_{\beta,\xi}\int \dd^2k\,
  (\varepsilon^\xi_\beta-\mu)\frac{\partial f}{\partial\varepsilon^\xi_\beta}\,\mathcal F^\xi_\beta(\mathbf k), \notag\\
\label{eq:conventional_nernst}
\\
\beta^{\mathrm A}_{xy}
&=&
-\frac{2eB_z}{\hbar T}
\sum_{\beta,\xi}
\int_{S_\xi}
\frac{\dd^2 k}{(2\pi)^2}
\left(
\varepsilon^\xi_\beta(\mathbf{k})-\mu
\right) \notag\\
& & \times
\frac{\partial f}
{\partial\varepsilon^\xi_\beta}
m^\xi_{z,\beta}(\mathbf{k})
\Omega^\xi_{z,\beta}(\mathbf{k}).
\label{eq:anomalous_nernst_linear_B}
\end{eqnarray}
\end{subequations}

Here, the additional factor $(\varepsilon-\mu)/T$ makes the thermoelectric response particularly sensitive to rapid energy variation of the corresponding electrical conductivity.
Similarly, the total transverse thermoelectric conductivity is given by the sum of its conventional and anomalous contributions, $\beta_{xy}^{\perp}=\beta_{xy}^{\text{cl} }+\beta_{xy}^{\mathrm A}$.

\section{Results}\label{sec:result}
\subsection{Displacement Field Tuning of Dirac-like Band Edges and Bilayer-like Dispersion Branch}

\begin{figure*}[t]
    \centering
    \includegraphics[width=\linewidth]{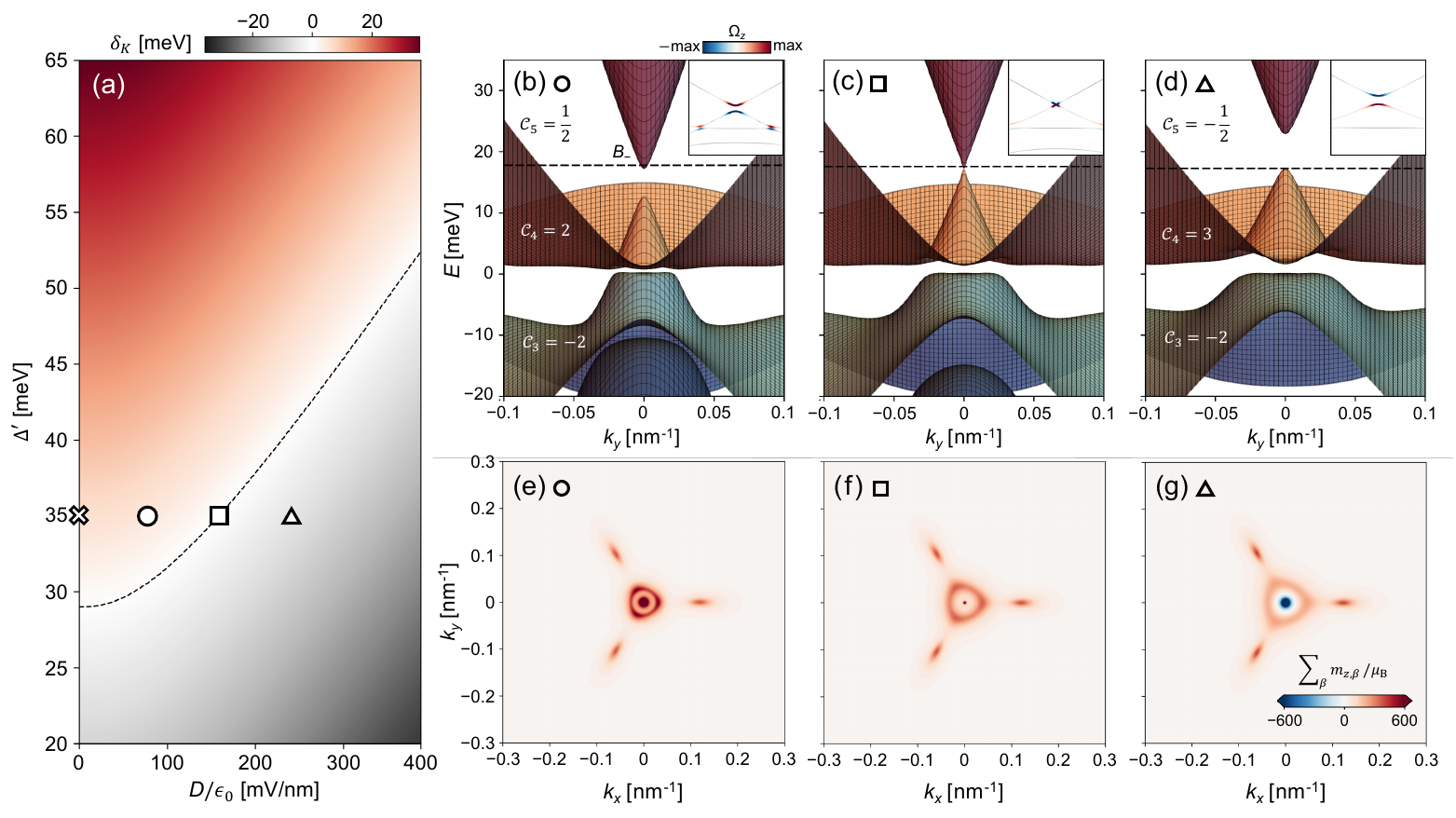}
    \caption{Band inversion in ABA TLG at the $\mathbf{K}$ valley. (a) Valley-centre gap $\delta_K$ as a function of displacement field $D/\epsilon_0$ and on-site energy parameter $\Delta^\prime$. The dashed line marks the analytical band-touching condition $\delta_K=0$. (b-d) Low-energy band structures, calculated with screening potentials at a carrier density of $n=7.34\times 10^{11}\text{cm}^{-2}$, at the marked parameter points, with insets showing the projected Berry curvature $\Omega_z$ onto the band structure. Crossing the dashed line inverts the ordering of $A_{up}$ and $B_-$-like states and the associated Berry curvature and orbital magnetic moment, with the valley Chern numbers of the relevant bands exchanging a unit of one across the inversion. (e-g) All-band summed orbital magnetic moment maps for the system. The largest moments are found in the fifth band, with peak magnitudes of the local valley g-factor, defined as $g_v=2m_z/\mu_B$, reaching approximately 5000.}
    \label{fig:band_closure}
\end{figure*}

\begin{figure*}[t]
    \centering    \includegraphics[width=0.95\linewidth]{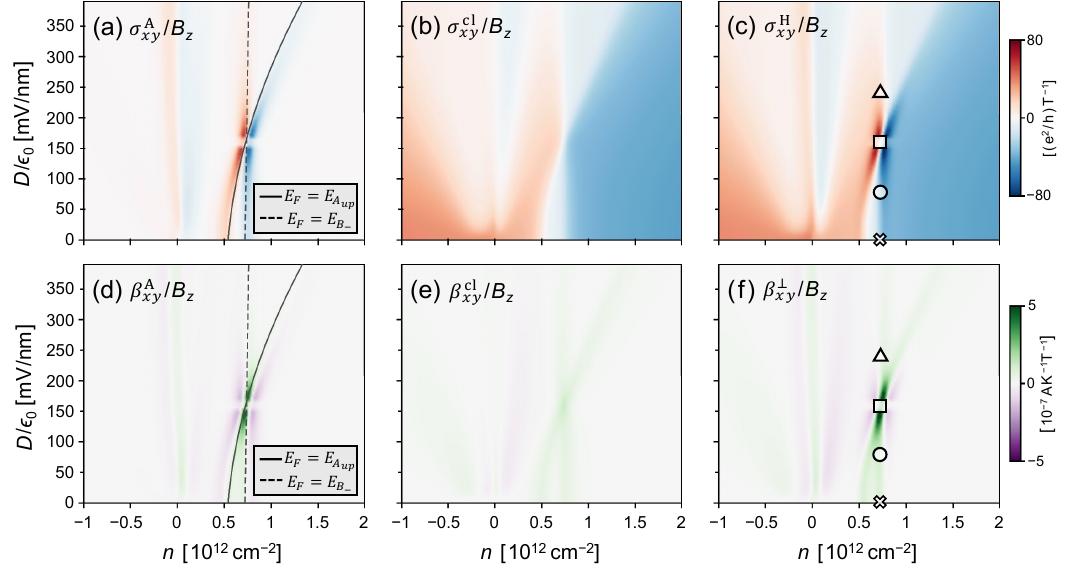}
    \caption{Weak field transverse transport response of ABA TLG for $\Delta^\prime=35$ meV and $T=10$ K. (a-c) Anomalous, semiclassical and total Hall responses $\sigma^\text{A}_{xy}/B_z$, $\sigma^\text{cl}_{xy}/B_z$, and $\sigma^\text{H}_{xy}/B_z$ respectively, calculated with a relaxation time $\tau=10\times10^{-14}$~s as functions of carrier density $n$ and displacement field $D/\epsilon_0$. (d-f) Corresponding anomalous, semiclassical and total transverse thermoelectric responses $\beta^\text{A}_{xy}/B_z$, $\beta^\text{cl}_{xy}/B_z$, and $\beta^\perp_{xy}/B_z$ respectively. The anomalous response is concentrated near the band-inversion region, while the semiclassical contribution extends over a broader range of carrier densities. The four markers correspond to the same parameter points as those marked in Fig.~\ref{fig:band_closure}(a).}
    \label{fig:trans_35_10}
\end{figure*}

We now consider the effect of mirror-symmetry breaking on the low-energy band structure of ABA TLG. As discussed in Sec.~\ref{sec:model}, a finite antisymmetric layer potential $\Delta_a (D)$ hybridises the monolayer- and bilayer-like branches \cite{koshino_gate-induced_2009,morimoto_gate-induced_2013}. For $\Delta_a\ll\gamma_1$, this coupling can be treated perturbatively around the mirror-symmetric limit. The effect of this hybridisation is particularly transparent at the valley centre ($\mathbf{k}=0$), where the relevant low-energy states are the upper $A$-branch eigenstate and the predominantly $B_-$-like state. We obtain
\begin{align}
    E_{A_{\text{up}}}(\mathbf{k}=0) &= 
    \Delta_s + \sqrt{\Delta_a^2 + \frac{\gamma_2^2}{4}}, \notag\\
    E_{B_-}(\mathbf{k}=0) 
    &=
    \Delta^\prime + \Delta_s - \frac{\gamma_5}{2}. \notag
\end{align}
Here, $E_{A_{\text{up}}}$ is obtained by exactly diagonalising the $A_{\pm}$ subspace, whereas $E_{B_-}$ is obtained perturbatively by neglecting higher-order corrections in $\Delta_a/\gamma_1$ [also see details in Appendix~\ref{appendix:perturb theory}]. The energy difference between these two states defines the  gap at the $\mathbf K$ valley,
\begin{equation}
\delta_K 
= E_{B_-}-E_{A_{\text{up}}} 
= \Delta' - \frac{\gamma_5}{2} - \sqrt{\big[\Delta_a (D)\big]^2+\frac{\gamma_2^2}{4}},
\label{eq:deltaK}
\end{equation}
where the displacement field dependence $\Delta_a(D)$ is given in Fig.~\ref{fig:screening}. From this we see that the symmetric potential $\Delta_s$ cancels from the  gap, while the mirror-breaking potential $\Delta_a(D)$ directly controls the band edge splitting. 

We also notice that, in principle, the gap may close and reopen with the opposite sign (indicating band inversion) if $\Delta_a$ is substantially increased, but only if the input value of $\Delta^\prime$ is large enough: $\Delta^\prime >(\gamma_5+|\gamma_2|)/2=29$ meV, which is estimated from SWMcC parameters in Ref.~\cite{morimoto_gate-induced_2013}. This possibility is illustrated in Fig.~ \ref{fig:band_closure}, where we identify conditions for such an inversion of Dirac-like band edges. In Fig.~\ref{fig:band_closure}(a) the dashed line marks the collapse of the band gap, $\delta_K=0$. Note that the dashed line starts at $\Delta'_c=(\gamma_5+|\gamma_2|)/2=29$ meV.

For $\Delta^\prime<29$ meV, the gap $\delta_K$ remains negative for any $\Delta_a$, and no field-driven band inversion occurs at the valley centre. We also note that the $\Delta^\prime$ parameter is not among those clearly established for bulk and thin-films of graphite, so that the experimental observation of the Dirac-like band inversion could be used for measuring its value, if $\Delta^\prime$ is large enough.

In some more detail, Fig.~\ref{fig:band_closure} panels (b–d) show the low-energy band structures at three points marked by a circle, square and triangle on the parametric map in Fig.~\ref{fig:band_closure}(a). As the displacement field increases, the gap between the fourth and fifth bands closes and reopens at the valley centre, reversing the ordering of the $A_{\text{up}}$ and $B_-$-like states. This band inversion is accompanied by a reversal of the Berry curvature near the valley centre and a change in the valley Chern numbers of the two bands by one, as indicated in Figs. \ref{fig:band_closure}(b) and \ref{fig:band_closure}(d).

The bilayer-like branch exhibits a distinct dependence on the electrostatic potentials. In the mirror-symmetric limit, the valley-centre separation between the low-energy $A_+$ and $B_2$ states is $\Delta_s+\gamma_2/2$. Upon breaking mirror symmetry, the lower $A_\pm$-derived level shifts to $E_{A_{\text{lo}}}=\Delta_s-\sqrt{\Delta_a^2+\gamma_2^2/4}$, while the $B_2$ level remains unchanged at the valley centre. This behaviour is distinct from the inversion of the Dirac-like branch considered above. For $|\Delta_a|\ll|\gamma_2|/2$, the valley-centre separation between the upper $A_\pm$-derived level and $B_-$ decreases by approximately $\Delta_a^2/|\gamma_2|$, bringing them closer to the discussed band inversion, whereas the separation between $B_2$ and $E_{A_{\text{lo}}}$ increases by approximately the same amount. Similar to the case for bilayer graphene \cite{kechedzhi2007trigonal,mccann2007lowenergy,koshino2007diamagnetism}, away from the valley centre, trigonal warping of the bilayer-like dispersion produces three symmetry-related satellite features at characteristic radial momentum $|k_\star| \simeq 2 \gamma_1 v_3 / (\hbar v^2) \approx 0.12~\text{nm}^{-1}$ \cite{koshino_landau_2011}. This is consistent with the three localised regions of enhanced positive orbital magnetic moment shown in Figs.~\ref{fig:band_closure}(e-g). At larger displacement fields, trigonal warping and interbranch hybridisation can additionally produce off-centre “Dirac gullies" in ABA TLG \cite{morimoto_gate-induced_2013,serbyn_new_2013,zibrov_emergent_2018,winterer_spontaneous_2022}.

%----------------------------------------------------------
\subsection{Carrier Density and Displacement Field Dependence of Hall and Transverse Thermoelectric Response}

Figure~\ref{fig:trans_35_10} shows the calculated transverse transport responses as functions of carrier density and displacement field for $\Delta'=35$ meV. The solid and dashed lines mark the conditions $E_F=E_{A_{\text{up}}}$ and $E_F=E_{B_-}$, respectively, indicating where the Fermi level crosses the corresponding valley-centre band edges.

The intersection of these  contours occurs for $E_{A_{\text{up}}}=E_{B_-}=E_F$ identifying the band touching at the Fermi level, corresponding to the inversion condition $\delta_K=0$ derived in Eq.~\eqref{eq:deltaK} occurring at the critical displacement field identified in Fig.~\ref{fig:band_closure} for $\Delta^\prime=35$~meV. The strongest anomalous Hall features therefore appear along these contours, where the Fermi level samples a strongly concentrated Berry curvature and orbital magnetic moment. The transport response therefore reflects the distribution of Berry curvature and also the occupation of the corresponding low-energy states, whose shift is determined by the magnetic moment.

As shown in Fig.~\ref{fig:trans_35_10}(a), the anomalous Hall response exhibits sharp features near the relevant band edges, where the product $m_z\Omega_z$ is strongly enhanced. 
In contrast, the conventional Hall response in Fig.~\ref{fig:trans_35_10}(b), governed by the Fermi-surface velocities and band curvatures, exhibits a pronounced sign change with carrier density and a distinct dependence on the displacement field.  
The total Hall response in Fig.~\ref{fig:trans_35_10}(c) combines these two contributions, with the sharp anomalous features superimposed on the conventional Hall background.
The features in the anomalous Hall effect are hence markedly distinct from a system lacking this band-inversion such as ABC TLG \cite{slizovskiy_films_2019}.

\begin{figure}[t]
    \centering    \includegraphics[width=0.75\linewidth]{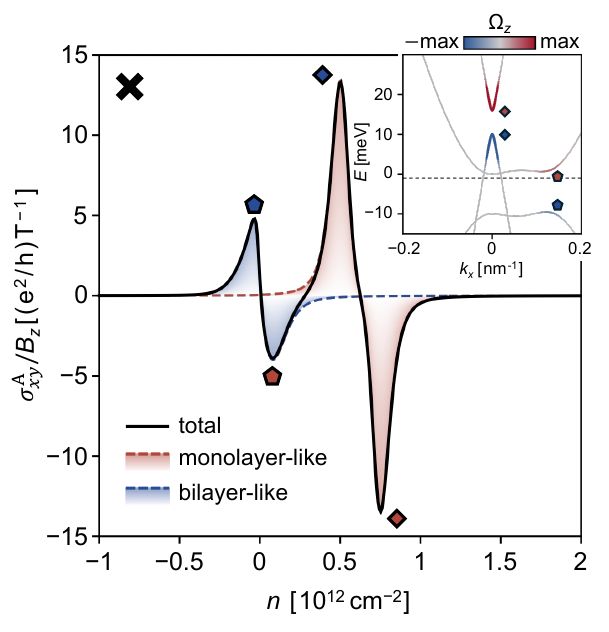}
    \caption{
    The anomalous Hall effect in mirror-symmetric ABA trilayer graphene. The Berry-curvature-weighted band structure shown in the inset illustrates the distinct contributions of the monolayer-like and bilayer-like branches.
    }
    \label{fig:mirror_geometry}
\end{figure}

The zero-displacement-field case, indicated by the cross in Figs.~\ref{fig:band_closure}(a), \ref{fig:trans_35_10}(c), and \ref{fig:trans_35_10}(f), is further examined in Fig.~\ref{fig:mirror_geometry}. In this mirror-symmetric limit ($\Delta_a=0$), the monolayer-like and bilayer-like branches are decoupled and contribute independently to the Berry curvature $\Omega_z$ and orbital magnetic moment $m_z$, allowing their respective contributions to the anomalous Hall conductivity to be distinguished. As shown in Fig.~\ref{fig:mirror_geometry}, the bilayer-like branch accounts for the more pronounced peaks in the total anomalous Hall response.

\begin{figure}[t]
    \centering
    \includegraphics[width=\linewidth]{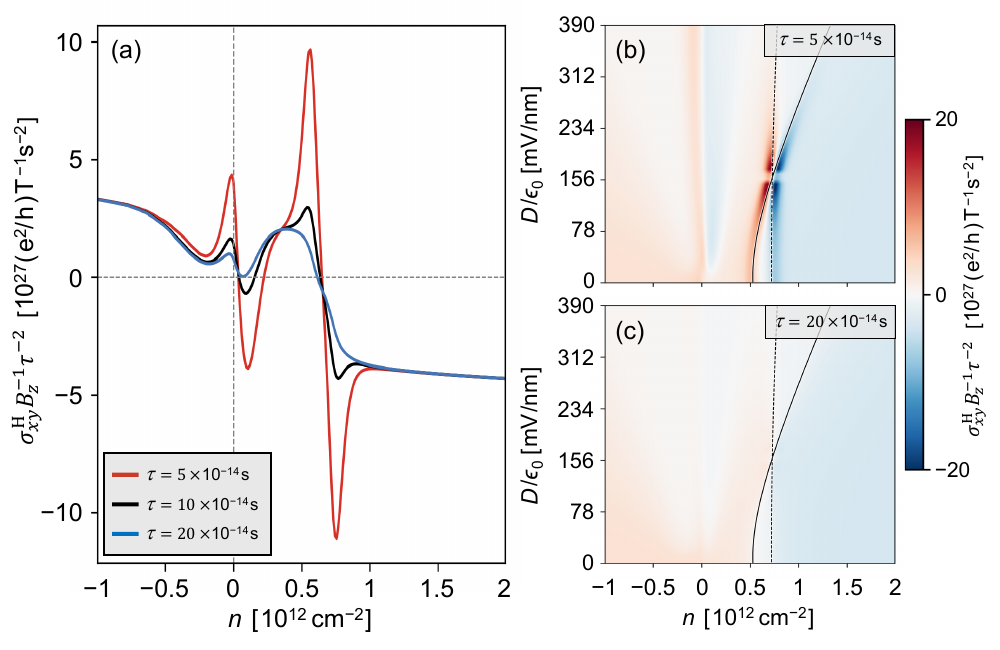}
    \caption{
    Dependence of the $\tau$-adjusted total Hall response on the relaxation time $\tau$ for $\Delta^\prime=35$~meV. (a) Carrier-density line cuts at $ D/\epsilon_0\approx78$ mV/nm (bare electrostatic potential 10 meV) for $\tau=(5,10,20)\times10^{-14}$ s. The anomalous contribution is independent of $\tau$, while the semiclassical Hall response scales as $\tau^2$, causing the inversion-related features to become less pronounced relative to the conventional background as $\tau$ increases. (b-c) Total Hall response in the $(n,D)$ plane for $\tau=5\times10^{-14}$s and $20\times10^{-14}$s respectively.}
    \label{fig:transport_linecut}
\end{figure}

The corresponding transverse thermoelectric responses are shown in Figs.~\ref{fig:trans_35_10}(d–f). The additional energy weighting $(\varepsilon-\mu)/T$ makes the thermoelectric response particularly sensitive to regions where the corresponding electrical conductivity varies rapidly with chemical potential, while suppressing slowly varying contributions. As shown in Fig.~\ref{fig:trans_35_10}(d), the anomalous thermoelectric response exhibits sharp sign-changing features near the relevant band edges, whereas the semiclassical contribution in Fig.~\ref{fig:trans_35_10}(e) is more broadly distributed. These sharp anomalous features remain visible in the total thermoelectric response in Fig.~\ref{fig:trans_35_10}(f).

Although the anomalous Hall response exhibits pronounced features near the relevant band edges, their prominence in the total Hall conductivity $\sigma^{\text{H}}_{xy}$ depends on the relative strength of the anomalous and semiclassical contributions. These two contributions exhibit different dependences on the relaxation time $\tau$. Within the constant-$\tau$ approximation, the semiclassical contribution scales as $\tau^2$, whereas the anomalous contribution is independent of $\tau$. To illustrate this effect, Fig.~\ref{fig:transport_linecut}(a) shows the total Hall conductivity normalised by $B_z\tau^2$ as a function of carrier density for three different relaxation times at a fixed displacement field. As $\tau$ increases, the sharp features associated with the band inversion become less pronounced relative to the semiclassical contribution. This behaviour is further illustrated by the transport maps in Figs.~\ref{fig:transport_linecut}(b,c), where the inversion-related features are more distinguishable at shorter relaxation times.

%----------------------------------------------------------
\section{Conclusion}\label{sec:conclusion}

In summary, our findings for electric-field tunable ABA trilayer graphene are:
\begin{itemize}[
    leftmargin=0pt,
    itemindent=1.2em,
    labelwidth=0.4em,
    labelsep=0.8em,
    itemsep=0.6em
]
    \item Depending on the values of SWMcC parameters $\gamma_2,\gamma_5$, and $\Delta^\prime$, the displacement field can induce a band inversion for the light branch of electrons, as described by Equation~\ref{eq:deltaK}.

    \item For Dirac-like light electrons, the band-edge inversion is accompanied by a reversal of the topological characteristics (Fig.~\ref{fig:band_closure}) and by a giant valley $g$-factor of the band-edge states.

    \item The resulting magnetic-field-induced valley splitting gives rise to finite anomalous transverse responses, with pronounced features near the low-energy band edges, in contrast to the more broadly distributed conventional semiclassical contributions, as shown in Fig.~\ref{fig:trans_35_10}.
\end{itemize}    
    
Potentially, detecting the band edge inversion for the Dirac-like trilayer bands (Eq.~\eqref{eq:deltaK}) may provide a route to measuring the value of uncertain SWMcC model parameters, such as the on-site energy potential $\Delta'$ and the hopping parameters $\gamma_2$ and $\gamma_5$.

%----------------------------------------------------------
\begin{acknowledgments}
A. M. and E. R. H. acknowledge support from the EPSRC 2D Materials of Tomorrow (2DMoT) CDT.
\end{acknowledgments}
%----------------------------------------------------------
\section*{Data availability}
The data that support the findings of this study are available from the authors upon reasonable request.
%----------------------------------------------------------
\appendix
%----------------------------------------------------------
\section{Detailed Derivation of the $\mathbf K$-Point Perturbation Theory}\label{appendix:perturb theory}

At the valley centre $\mathbf{k}=0$, all momentum-dependent terms in Eq.~\eqref{Hamiltonian} vanish. In the mirror-symmetric basis introduced in Sec.~\ref{subsec:mirror sym}, the dimer pair $(B_+,A_2)$ is coupled by $\sqrt{2}\gamma_1$. We rotate it to $B_+^{\mathrm{rot}}=\cos\theta\,B_+ + \sin\theta\,A_2$ and $A_2^{\mathrm{rot}}=\sin\theta\,B_+ - \cos\theta\,A_2$, with
$\tan 2\theta = 2\sqrt{2}\gamma_1/(\gamma_5/2-\Delta')$, which diagonalises this block at $\Delta_s=0$ with eigenvalues
$E_{B_+},E_{A_2} = \tfrac{1}{2}(3\Delta'+\gamma_5/2) \pm \sqrt{(\gamma_5/2-\Delta')^2/4 + 2\gamma_1^2}$.  In the reordered basis $(A_-,A_+,B_-,B_+^{\mathrm{rot}},A_2^{\mathrm{rot}},B_2)$ the Hamiltonian is block diagonal:
\begin{widetext}
    \begin{equation}
        \left(
    \begin{array}{cccccc}
     \Delta_s-\frac{\gamma_2}{2} & -\Delta_a & 0 & 0 & 0 & 0 \\
     -\Delta_a & \Delta_s+\frac{\gamma_2}{2} & 0 & 0 & 0 & 0 \\
     0 & 0 & \Delta^\prime+\Delta_s-\frac{\gamma_5}{2} & -\Delta_a\cos\theta & -\Delta_a\sin\theta & 0 \\
     0 & 0 & -\Delta_a\cos\theta &
     E_{B_+}+\Delta_s\cos^2\theta
     &
     \Delta_s\cos\theta\sin\theta & 0 \\
     0 & 0 & -\Delta_a\sin\theta &
     \Delta_s\cos\theta\sin\theta
     &
     E_{A_2}+\Delta_s\sin^2\theta & 0 \\
     0 & 0 & 0 & 0 & 0 & 0 \\
    \end{array}
    \right).
    \end{equation}    
\end{widetext}

The $2\times 2$ upper-left block $H_2$ involves only the $A_-$ and $A_+$ states, which can be solved exactly to give the eigenvalues 
\begin{align}
E_{A_{\text{up/lo}}} = \Delta_s \pm \sqrt{\Delta_a^2 + \frac{\gamma_2^2}{4}} \approx \Delta_s \pm \frac{|\gamma_2|}{2} \left(1 + \frac{2\Delta_a^2}{\gamma_2^2}\right),   
\end{align}
with the approximate form valid for $|\Delta_a|\ll|\gamma_2|/2$.

The $3\times3$ block $H_3$, spanned by $B_-$, $B_+^{\mathrm{rot}}$ and $A_2^{\mathrm{rot}}$, couples $B_-$ to the dimer states only through terms proportional to $\Delta_a$. We treat both $\Delta_a$ and $\Delta_s$ as perturbations, retaining terms up to second order. The $B_2$ state remains decoupled at
$\mathbf{k}=0$, with zero energy. Applying second-order perturbation theory to the $B_-$-like state, we obtain the energy expression:
\begin{equation}
    E_{B_-} = 
    \Delta^\prime + \Delta_s
    - \frac{\gamma_5}{2}
    + \frac{\Delta_a^2  \left( 2\Delta' + \gamma_5  \right)}
    {4\gamma_1^2 - \gamma_5 \left( 2\Delta' + \gamma_5  \right)}.
\end{equation}
Since the primary interlayer hopping $\gamma_1 \approx 390~\text{meV}$ is the dominant energy scale in the system, the second-order correction term is of order $\mathcal{O}(\gamma_1^{-2})$ and contributes less than $1\text{ meV}$ under realistic field strengths. For example, at $\Delta_a=30~\text{meV}$ and $\Delta^\prime=35~\text{meV}$ the correction is $\approx 0.16~\text{meV}$. Neglecting this higher-order term yields the simplified $E_{B_-}$ formula used in the main text. Finally, subtracting $E_{A_{\rm up}}$ from $E_{B_-}$ gives Eq.~\eqref{eq:deltaK}. Setting $\delta_K=0$ at $\Delta_a=0$
gives the threshold $\Delta'_c=(\gamma_5+|\gamma_2|)/2$, above which $\delta_K$ changes sign at finite $\Delta_a$.

%----------------------------------------------------------
\section{A Comparison of the Transport Response Maps Under Different Parameter Sets}
\begin{figure}
    \centering
    \includegraphics[width=\linewidth]{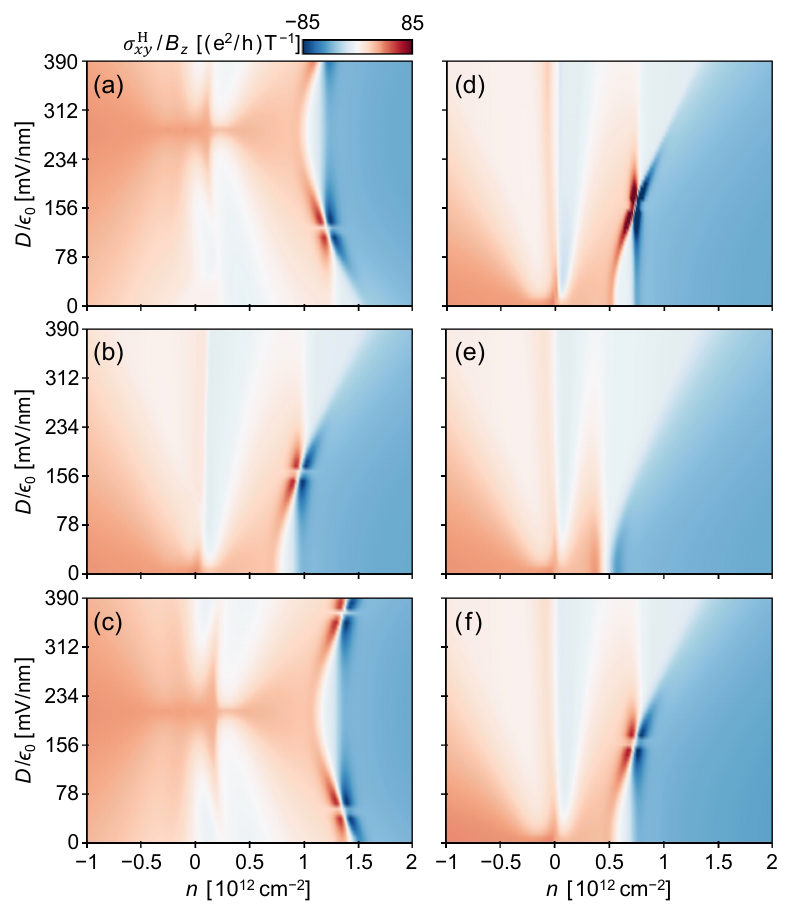}
    \caption{(a)–(f) Theoretically calculated total Hall conductivity per unit magnetic field, $\sigma_{xy}^{\text{H}}/B_z$, under various parameter selections, whose values are summarised in Table~\ref{tab:comparison}. }
    \label{fig:comparison}
\end{figure}

\begin{table}[h]
    \centering
    \caption{A summary of parameter values used in Figs.~\ref{fig:comparison}(a)--\ref{fig:comparison}(f).}
    \label{tab:comparison}
    \renewcommand{\arraystretch}{1.2}
    \begin{tabular}{ccccc}
        \toprule
        Figure 
        & $\Delta_t$ [meV]
        & $\Delta_b$ [meV] 
        & $\Delta'$ [meV]
        & $T$ [K] \\
        \midrule
        \ref{fig:comparison}(a) & 0  & 70 & 35 & 10 \\
        \ref{fig:comparison}(b) & 18 & 18 & 35 & 10 \\
        \ref{fig:comparison}(c) & 18 & 70 & 35 & 10 \\
        \ref{fig:comparison}(d) & 0  & 0  & 35 & 5  \\
        \ref{fig:comparison}(e) & 0  & 0  & 25 & 10 \\
        \ref{fig:trans_35_10}(c) and \ref{fig:comparison}(f) & 0  & 0  & 35 & 10 \\
        \bottomrule
    \end{tabular}
\end{table}

Our theoretical model involves a series of system parameters whose values can quantitatively change the calculated
transport coefficients. 
$\Delta_t$ and $\Delta_b$ account for the influence of the dielectric environment, as TLG is usually surrounded by dielectric materials such as hBN and SiO$_2$.
For a more comprehensive perspective, we vary these parameters and show the total Hall conductivity per unit magnetic field $\sigma^{\text{H}}_{xy}/B_z$ in Figs. \ref{fig:comparison}(a)-(f), with parameter values summarized in Table \ref{tab:comparison}.
As can be seen from Fig.~\ref{fig:comparison}, although the location and magnitude of the response features vary with dielectric offsets, temperature, and $\Delta^\prime$, the response associated with the band-inversion region of the $(D,n)$ parameter space persists across parameter sets, provided that $\Delta^\prime > (\gamma_5+|\gamma_2|)/2$. In contrast, panel (e), with $\Delta^\prime<\Delta^\prime_c$, shows no inversion feature.

%----------------------------------------------------------
\bibliography{references}

\end{document}